# Phase diagrams of a cylindrical spin-1 ising nanowire within the presence of the crystal field and the transverse field

Ersin Kantar[a,*] and Yusuf Kocakaplan[b]

[a]*Department of Physics, Erciyes University, 38039 Kayseri,Turkey*

[b]*Institute of Science, Erciyes University, 38039 Kayseri,Turkey*

## Abstract

Within the framework of the effective-field theory (EFT) with correlations, the phase diagrams of a cylindrical spin-1 Ising nanowire is investigated, depending on the ratio of the Hamiltonian parameters. We discuss in detail the influence of interfacial coupling, crystal field and transfer field ratio on the phase diagrams, and some characteristic phenomena have been found such as tricritical point and first-order transitions.

## 1. Introduction

In the past two decades, magnetic properties of the nanomaterials are quite different from those of the bulk and are greatly affected by the particle size [1–4]. Furthermore, due to their technological [5–8] and biomedical applications [9–13] are a subject of great interest for both experimental and theoretical researchers. On the other hand, magnetic properties of nanoparticles have been studied within various Ising systems consisting of core-shell structures via a variety of theoretical techniques. Wang et al. [14] investigated phase transitions in magnetic alloy nanoparticles by the variational cumulant expansion in spin-1/2 system. The spin-1/2 Ising system within Monte Carlo (MC) simulations studied by Inglesias and Labarta [15, 16] and also investigated by Crisan et al. [17]. Kaneyoshi used the spin-1/2 Ising system and studied phase diagrams [18], magnetizations [19], the possibility of a compensation point [20] and reentrant phenemona [21] of nanoparticles with a core/shell structure by using the effective-field theory (EFT) with correlations. The magnetizations and susceptibilities of cylindrical core/shell spin-1 Ising nanowire have been investigated within the EFT with correlations [22]. Within effective-field theory based on a probability distribution technique, Kocakaplan et al. [23] studied hysteresis loops and compensation behavior of cylindrical transverse spin-1 Ising nanowire with the crystal field. Canko et al. [24] investigated some characteristic behaviors of spin-1 Ising nanotube.The hysteresis behaviors [25], and phase diagrams and magnetic properties [26] of a spin-1 Ising nanotube, with ferro- or anti-ferromagnetic interfacial coupling are examined in the presence of a random magnetic field by Magoussi et al.. On the other hand, in a few study are used binary mixed spins Ising system to examine nanoparticles system, namely Zaim et al. [27] used the binary mixed spins (1/2, 1) Ising system by the use of MC Simulations, Zaim and Kerouad [28] studied binary mixed spins (1/2, 1 or 3/2) by using MC simulations. Yüksel et al. [29] applied Monte Carlo simulations for the mixed spins (3/2 - 1) system. Recently, Jiang et al. [30] studied spin–3/2 core and a hexagonal ring spin-5/2 shell by the EFT with self-spin correlations. Kantar et al. [31], within the framework of the EFT with correlations, investigated spin–1/2 core surrounded by spin -3/2 the 2D nanoparticles system. Kantar and Kocakaplan [32] presented the phase diagrams and compensation behaviors hexagonal type Ising nanowire with spin-1/2 core and spin-1 shell structure.

*Corresponding author.
Tel: + 90 352 207 66 66 # 33136; Fax: + 90 352 4374931.
E-mail address: ersinkantar@erciyes.edu.tr (E. Kantar)

In the present study, we investigated the phase diagrams in difference plane of the cylindrical core/shell spin-1 Ising nanowire within the EFT with correlations. The influence of interfacial coupling, crystal field and transfer field ratio on the phase diagrams have been illustrated.

The review of this paper is organized as follows. In Section 2, we give the model and present the formalism of the EFT with correlations. The detailed numerical results and discus-sions of phase diagrams are presented in Section 3. Finally Section 4 is devoted to a summary and a brief conclusion.

## 2. Model and formulation

We consider a cylindrical Ising nanowire, the schematic representation as depicted in Fig. 1 in which the wire consists of the surface shell and core, and each site is occupied by a spin-1 Ising atom on the figure. With an exchange interaction, each spin is connected to the two nearest- neighbor spins on the above and below sections along the cylinder. So, the Hamiltonian of the system can be expressed as

$$H=-J_S\sum_{\langle ij\rangle}S_i^zS_j^z-J_C\sum_{\langle mn\rangle}S_m^zS_n^z-J_{Int}\sum_{\langle im\rangle}S_i^zS_m^z-\Delta\left(\sum_i\left(S_i^z\right)^2+\sum_m\left(S_m^z\right)^2\right)$$
$$-\Omega_s\sum_i S_i^x-\Omega_c\sum_m S_m^x \qquad (1)$$

where $S_i^z$ and $S_i^x$ denote the z and x components of the quantum spin $\vec{S}$ operator, respectively. The $S_i^z$ and $S_i^x$ components take values $0, \pm 1$ at each site $i$. $J_s$, $J_c$ and $J_1$ are the exchange interaction parameters between two nearest-neighbor quantum spin operators in the surface shell, the core and between the surface shell and core, respectively. The first three summations are over the nearest-neighbor pairs of spins at the surface shell, the core and between the surface shell and core, respectively. $\Delta$, $\Omega_c$, $\Omega_s$ and $h$ Hamiltonian parameters display the single-ion anisotropy (i.e. crystal field), the transverse field in the core, the transverse field at the surface shell, and longitudinal magnetic field, respectively. In order to clear up the influences of surface on physical properties, the surface exchange interaction parameter is often defined as $J_s = J_c(1+\Delta_s)$ in the nanosystems.

Within the framework of the EFT with correlations, one can easily find the longitudinal (z-direction) magnetizations $m_{S1}$ and $m_{S2}$ at the surface shell and the longitudinal magnetizations $m_{C1}$ and $m_{C2}$ in the core, as coupled equations,

$$m_{S1} = \left[1+m_{S1}\,sinh(A)+m_{S1}^2\left(cosh(A)-1\right)\right]^2\left[1+m_{S2}\,sinh(A)+m_{S2}^2\left(cosh(A)-1\right)\right]^2 \left[1+m_{C2}\,sinh(B)+m_{C2}^2\left(cosh(B)-1\right)\right]F_S(x)\big|_{x=0}, \tag{2a}$$

$$m_{S2} = \left[1+m_{S2}\,sinh(A)+m_{S2}^2\left(cosh(A)-1\right)\right]^2\left[1+m_{S1}\,sinh(A)+m_{S1}^2\left(cosh(A)-1\right)\right]^2 \left[1+m_{C2}\,sinh(B)+m_{C2}^2\left(cosh(B)-1\right)\right]^2 F_S(x)\big|_{x=0}, \tag{2b}$$

$$m_{C1} = \left[1+m_{C1}\,sinh(C)+m_{C1}^2\left(cosh(C)-1\right)\right]^2\left[1+m_{C2}\,sinh(C)+m_{C2}^2\left(cosh(C)-1\right)\right]^6 F_C(x)\big|_{x=0}, \tag{2c}$$

$$m_{C2} = \left[1+m_{C2}\,sinh(C)+m_{C2}^2\left(cosh(C)-1\right)\right]^4\left[1+m_{C1}\,sinh(C)+m_{C1}^2\left(cosh(C)-1\right)\right] \left[1+m_{S1}\,sinh(B)+m_{S1}^2\left(cosh(B)-1\right)\right]\left[1+m_{S2}\,sinh(B)+m_{S2}^2\left(cosh(B)-1\right)\right]^2 F_C(x)\big|_{x=0}. \tag{2d}$$

In here, the term A, B and C defined as $A=J_S\nabla, B=J_1\nabla$ and $C=J_C\nabla$. $\nabla=\partial/\partial x$ is the differential operator. The functions $F_S(x)$ and $F_C(x)$ are defined in [23] as follow:

$$F_{S,C}(x)=\sum_{n=1}^{3}\left(exp\left[2\beta\gamma\,cos(\varphi)+\frac{2\beta\Delta}{3}\right]\right)\left(\frac{2x}{3\gamma}cos(\varphi)+\frac{2}{27}\frac{\Delta^3x-\Delta x^3+7/2\Delta\Omega^2x}{B\gamma}sin(\varphi)\right) \left(\sum_{n=1}^{3}exp\left[2\beta\gamma\,cos(\varphi)+\frac{2\beta\Delta}{3}\right]\right)^{-1} \tag{3}$$

where

$$\theta = arccos\left[\frac{A}{\gamma^3}\right],$$

$$A = -\frac{1}{27}\Delta^3 + \frac{1}{3}\Delta x^2 - \frac{1}{6}\Delta\Omega^3,$$

$$B = \frac{1}{9}\left(-\Delta^2 x^4 + \frac{3}{4}\Delta^2\Omega^4 - 3\Delta^4 x^2 + 15\Delta^2 x^2\Omega^2 + 3\left(x^2+\Omega^2\right)^3\right)^{1/2}, \tag{4}$$

$$\gamma = \left(A^2 + B^2\right)^{1/6}$$

and

$$\varphi = \frac{(n-1)2\pi + \theta}{3}.$$

It would be worthwhile to examine for two special cases: (i) the first case correspond to Blume-Capel model and in this case $\Omega = 0.0$. The second case correspond to spin-1 transverse Ising model and in this case Δ = 0.0. For the first case by using Eqs (3) and (4), we can reduce Eq. (3) to the form as seen in Eq. (5a) below. In similar, for the second case, $F_S\left(x\right)$ and $F_C\left(x\right)$ functions can be obtain in the form as seen in Eqs. (5b) below.

$$F_{S,\,C}\left(x\right) = \frac{2\sinh[\beta x]}{\exp(-\beta\Delta) + 2\cosh[\beta x]} \tag{5a}$$

$$F_{S,\,C}\left(x\right) = \frac{2x\sinh\left(\sqrt{\Omega_{S,C} + x^2}\beta\right)}{\left(\sqrt{\Omega_{S,C} + x^2}\right)\left(1 + 2\cosh\left(\sqrt{\Omega_{S,C} + x^2}\beta\right)\right)} \tag{5b}$$

Here, $\beta = 1/k_B T$, T is the absolute temperature and $k_B$ is the Boltzmann constant.

On the other hand, in order to obtain the phase diagram of the system, we must expand the right-hand sides of the (2a)-(2d) coupled equations. They are obtained as follows:

$$Km = \begin{pmatrix} a_1 & a_2 & 0 & 0 \\ b_1 & b_2 & b_3 & b_4 \\ 0 & c_1 & c_2 & c_3 \\ 0 & d_1 & d_2 & d_3 \end{pmatrix} \begin{pmatrix} m_{C1} \\ m_{C2} \\ m_{S1} \\ m_{S2} \end{pmatrix} = 0, \tag{6}$$

Here, the coefficients $a_i, b_i, c_i$ and $d_i$ in each matrix take complicated forms, so that they give in **Apendix A**. These coefficients can be easily obtained from the coupled equations via differential operator technique. The transition temperature of each system can be determined from det(K) = 0. We should also mention that, for the following discussions, we define the parameters *q* and *r* as

$$q = \frac{\Omega_S}{\Omega_C} \text{ and } r = \frac{J_{Int}}{J_C}. \tag{7}$$

## 3. Numerical results and discussions

In this section, we studied some interesting and typical results of the cylindrical core/shell spin-1 Ising nanowire with a crystal field and transverse field at a zero longitudinal magnetic field. The effects of the interfacial coupling, crystal field and transverse field on the phase diagrams of the cylindrical core/shell spin-1 Ising nanowire have investigated for selected values of the *r*, $\Delta_S$ , $\Omega_s$ , $\Omega_c$ and Δ.

### *3.1. Crystal field dependence of the Phase Diagrams*

Fig. 2 shows the phase diagram in the (T, Δ) plane for diverse values of the Hamiltonian parameters. In figures, the solid and dashed lines stand for the second- and first-order phase transition lines, respectively and the tricritical point (TCP) is illustrated by a black circle. All transitions are from ferromagnetic (f) to paramagnetic (p) phase. For the fixed values of $\Omega_c = 0.0$ and $\Delta_S = 0.0$ and different values of the interfacial coupling (r = 0.01, 0.5 and 1.0), Fig. 2(a) is plotted. We can clearly see that with the increase of r, the transition temperature of the system is goes up as well as the TCP in Fig. 2(a). The first-order phase transition temperature undergo zero roughly at the same value of Δ for r = 0.01 and 0.5 values. Fig. 2(b) is ob-tained for various values of surface exchange interaction parameter ($\Delta_S$ = -1.0, 0.0 and 1.0) and fixed values of $\Omega_c = 0.0$ and $r = 1.5$. The behaviours of Fig. 2(b) is similar to Fig. 2(a).

### *3.2. Core transverse field dependence of the Phase Diagrams*

Fig. 3 display the phase diagram in the (T, $\Omega_c$) plane. In Fig. 3, we can clearly see that all transitions are the second-order phase transitions and transitions are from ferromagnetic (f) to paramagnetic (p) phase. Fig. 3(a) is plotted for the fixed values of $r = 1.0$, $\Delta = 0.0$ and $\Omega_S = 1.0$ and different values of surface exchange interaction parameter ($\Delta_S$ = -1.0, 0.0 and 0.5). The transition temperature of the system is goes up with the increase of the $\Delta_S$. For the fixed values of $r = 1.0$, $\Delta = 0.0$, and $\Delta_S = 0.0$ and different values of the ratio of shell transverse field to the core transverse field (q = 1.0, 5.0 and 20.0), Fig. 3(b) is obtained. The second-order phase transition temperatures go to zero as $\Omega_c$ increase. On the other hand, Fig. 3(c) is plotted for the different values of interfacial interaction parameter ($r$ = 0.01, 0.5 and 1.0) and fixed values of $\Delta_S = 0.0$ and $\Delta = 0.0$. It is similar to the Fig. 3(b) except that at zero $\Omega_c$, transition temperature takes different values for selected values of r.

### *3.3. Transverse fields ratio dependence of the Phase Diagrams*

Fig. 4 illustrate the phase diagram in the (T, q) plane. Similar to Fig. 3, all transitions are the second-order phase transitions and transitions are from ferromagnetic (f) to paramagnetic (p) phase in Fig. 4. Fig. 4(a) is obtained for the diverse values of surface exchange interaction parameter ($\Delta_S$ = -1.0, 0.0 and 0.5) and fixed values of $\Omega_C = 1.0, \Delta = 0.0$ and $r = 1.0$. With the increase of the $\Delta_S$, the transition temperature of the system is raise. Also, when transverse fields ratio are grow transition temperature close to each other for selected different values of $\Delta_S$. For the fixed values of $r = 1.0$, $\Delta = 0.0$, and $\Delta_S = 0.0$ and different values of the ratio of core transverse field ($\Omega_C$ = 1.0, 3.0 and 5.0), Fig. 4(b) is obtained. As the value of $\Omega_C$ increase, the transition temperature value is go down. Finally, Fig. 4(c) is plotted for for the change values of interfacial interaction parameter ( $r = 0.01$, 0.5 and 1.0) and fixed values of $\Delta_S = 0.0$, $\Omega_C = 2.0$ and $\Delta = 0.0$. The remarkable feature of the Fig. 4(c) is that for the small value of r, the transition temperature is almost to be fixed as the increase of q. We should also mention that when q ratio is increase transition temperature close to each other for selected different values of r.

In summary, we have studied the phase diagrams of the cylindrical core/shell spin-1 Ising nanowire with a crystal field transverse field at zero longitudinal magnetic field via the EFT with correlation. We have obtained the phase diagrams of the system on the three different plane, namely (T, Δ), (T, $\Omega_c$) and (T, q). The effect of the Hamiltonian parameters on phase diagrams are explained in detail. We observed the system undergoes first- and second-order phase transitions as well as the TCP for the certain values of Hamiltonian parameters. Finally, we hope that our detailed theoretical investigations may stimulate further researches to study magnetic properties of nanoparticles systems, and also will motivate experimentalists to investigate the behaviors in real nanomaterial systems.

**Appendix A.**

$a_1$ = -$F_C$ [-$J_C$] +$F_C$ [$J_C$]; $a_2$ = -3$F_C$ [-$J_C$] +3$F_C$ [$J_C$],

$b_1$ = -(1/2) $F_C$ [-$J_C$] +1/2 $F_C$ [$J_C$]; $b_2$ = -2 $F_C$ [-$J_C$] +2 $F_C$ [$J_C$]; $b_3$ = -(1/2) $F_C$ [-$J_1$] +1/2 $F_C$ [$J_1$]; $b_4$ = -$F_C$ [-$J_1$] +$F_C$ [$J_1$],

$c_1$ = -(1/2) $F_S$ [-$J_1$] +1/2 FS [$J_1$]; $c_2$ = -$F_S$ [-$J_S$] +$F_S$ [$J_S$]; $c_3$ = -$F_S$ [-$J_S$] +$F_S$ [$J_S$],

$d_1$ = -$F_S$ [-$J_1$] +$F_S$ [$J_1$]; $d_2$ = -$F_S$ [-$J_S$] +$F_S$ [$J_S$]; $d_3$ = -$F_S$ [-$J_S$] +$F_S$ [$J_S$].

**References**

1. A.E. Berkowitz, R.H. Kodama, Salah A. et al., J. Magn. Magn. Mater. 196, 591 (1999).

2. V. Skumryev, S. Stoyanov, Y. Zhang, et al., Nature 423, 850 (2003).

3. C. Frandsen, C.W. Ostenfeld, M. Xu,et al., Phys. Rev. B 70, 134416 (2004).

4. X. He, Z.H. Wang, D.Y. Geng, et al., J. Mater. Sci. Tech. 27, 503 (2011).

5. R.H. Kadoma, J. Magn. Magn. Mater. 200, 1601 (1999).

6. C.J. O'Connor, J. Tang, and H. Zhang, in: J.S. Miller, M. Drillon, (Eds.), Magnetism: Molecules to Materials III, Wiley-VCH, Weinheim, (2002).

7. Q.A. Pankhurst, N.K.T. Thanh, S.K. Jones, et al., J. Phys. D: Appl. Phys. 42, 224001 (2009).

8. Q. Dai, D. Berman, K. Virwani, et al., Nano Lett. 10, 3216 (2010).

9. Q.A. Pankhurst, J. Connolly, S.K. Jones, et al., J. Phys. D: Appl. Phys. 36, R167 (2003).

10. S. Laurent, D. Forge, M. Port, et al., Chem. Rev. 108, 2064 (2008).

11. C.G. Hadjipanayis, M.J. Bonder, S. Balakrishnan, et al., Small 4, 1925 (2008).

12. N. Sounderya, Y. Zhang, Recent Pat. Biomed. Eng. 1, 34 (2008).

13. J. Rivas, M. Bañobre-López, Y. Piñeiro-Redondo, et al., J. Magn. Magn. Mater. 324, 3443 (2012).

14. H. Wang, Y. Zhou, D.L. Lin, et al., Phys. Status Solidi B 232, 254 (2002).

15. Ó. Iglesias, A. Labarta, Phys. Rev. B 63, 184416 (2001).

16. Ó. Iglesias, A. Labarta, Physica B 343, 286 (2004).

17. O. Crisan, E.E. Tornau, V. Petrauska, et al., Phys. Status Solidi C 1, 3760 (2004).

18. T. Kaneyhoshi, Phys. Status Solidi B 242, 2938 (2005).

19. T. Kaneyhoshi, J. Magn. Magn. Mater. 321, 3430 (2009).

20. T. Kaneyhoshi, Solid State Commun. 152, 883 (2012).

21. T. Kaneyhoshi, J. Magn. Magn. Mater. 339, 151 (2013).

22. N. Şarlı, M. Keskin, Solid State Commun. 152, 354 (2012).

23. Y. Kocakaplan, E. Kantar, and M. Keskin, European Physical Journal B 86, 420 (2013).

24. O. Canko, A. Erdinç , F. Taşkin, et al., Phys. Lett. A 375, 3547 (2011).

25. H. Magoussi, A. Zaim, and M. Kerouad, Chinese Physics B 22, 116401 (2013).

26. H. Magoussi, A. Zaim, and M. Kerouad J. Magn. Magn. Mater. 344, 109 (2013).

27. A. Zaim, M. Kerouad, and Y. EL Amraoui, Physica A 321, 1077 (2009).

28. A. Zaim, M. Kerouad, Physica A 389, 3435 (2010).

29. H. Magoussi, A. Zaim, and M. Kerouad, J. Magn. Magn. Mater. 344, 109 (2013).

30. A. Zaim, M. Kerouad, and M. Boughrara, Solid State Commun. 158, 76 (2013).

31. Y. Yüksel, E. Aydıner, and H. Polat, J. Magn. Magn. Mater. 323, 3168 (2011).

32. E. Kantar, Y. Kocakaplan, Solid State Communications, 177, 1 (2014).

**List of the figure captions**

**Fig. 1.** (Color online) Schematic presentation of a cylindrical core/shell spin-1 Ising nanowire. (a) Cross-section and (b) three-dimensional. The grey and blue circles illustrate magnetic atoms at the surface shell and core, respectively.

**Fig. 2.** (Color online) The phase diagrams in (T, $\Delta$) plane of the cylindrical core/shell spin-1 Ising nanowire. Dashed and solid lines represent the first- and second-order phase transitions, respectively. The tricritical points are indicated with black circles.

**(a)** For the fixed value of $\Omega_c = 0.0$, $\Delta_S = 0.0$ and r = 0.01, 0.5 and 1.0.

**(b)** For the fixed value of $\Omega_c = 0.0$, $r = 1.5$ and $\Delta_S$ = -1.0, 0.0 and 1.0.

**Fig. 3.** (Color online) The phase diagrams in (T, $\Omega_c$) plane.

**(a)** For the fixed value of r = 1.0, $\Delta = 0.0$ and $\Delta_S$ = -1.0, 0.0 and 0.5.

**(b)** For the fixed value of r = 1.0, $\Delta = 0.0$ and $q$ = 1.0, 5.0 and 20.0.

**(c)** For the fixed value of $\Delta_S = 0.0$, $\Delta = 0.0$ and r = 0.01, 0.5 and 1.0.

**Fig. 4.** (Color online) The phase diagrams in (T, q) plane.

**(a)** For the fixed value of r = 1.0, $\Omega_c = 1.0$, $\Delta = 0.0$ and $\Delta_S$ = -1.0, 0.0 and 0.5.

**(b)** For the fixed value of r = 1.0, $\Delta_S = 0.0$, $\Delta = 0.0$ and $\Omega_c$ = 1.0, 3.0 and 5.0.

**(c)** For the fixed value of $\Delta_S = 0.0$, $\Omega_c = 2.0$, $\Delta = 0.0$ and r = 0.01, 0.5 and 1.0.

**Figure 1**

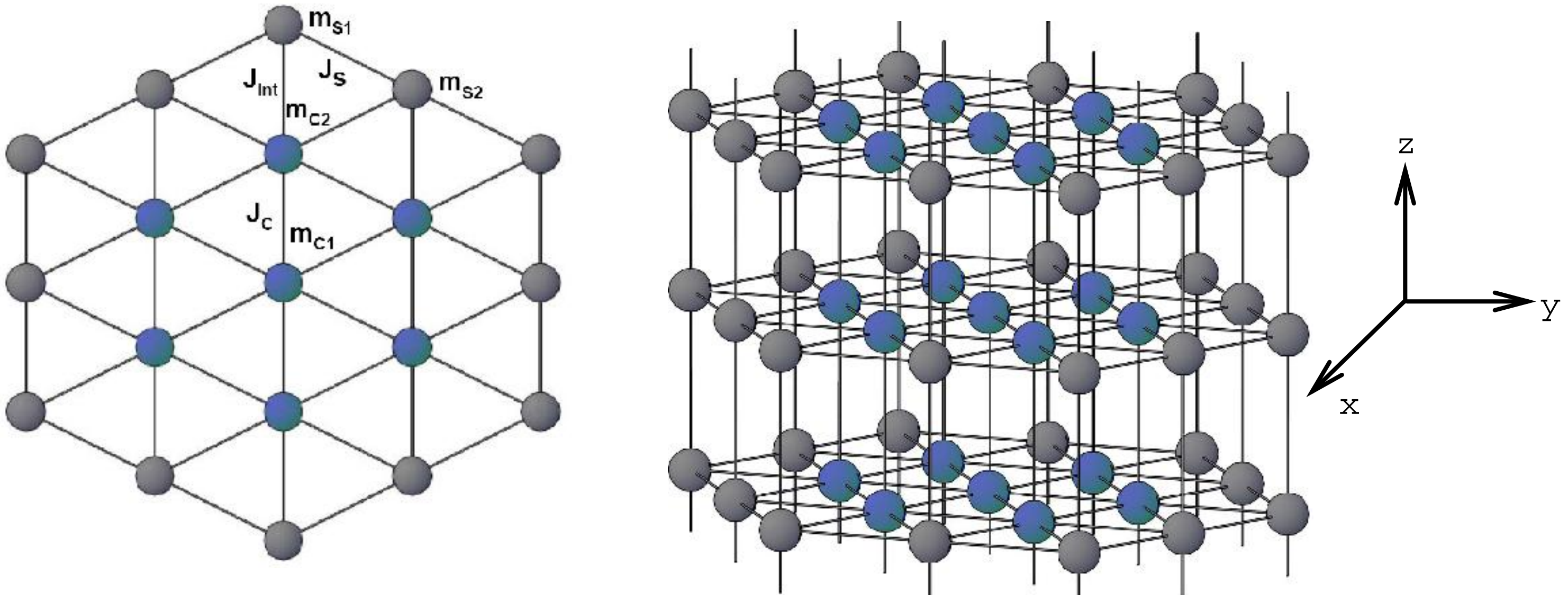

(a) (b)

Figure 2

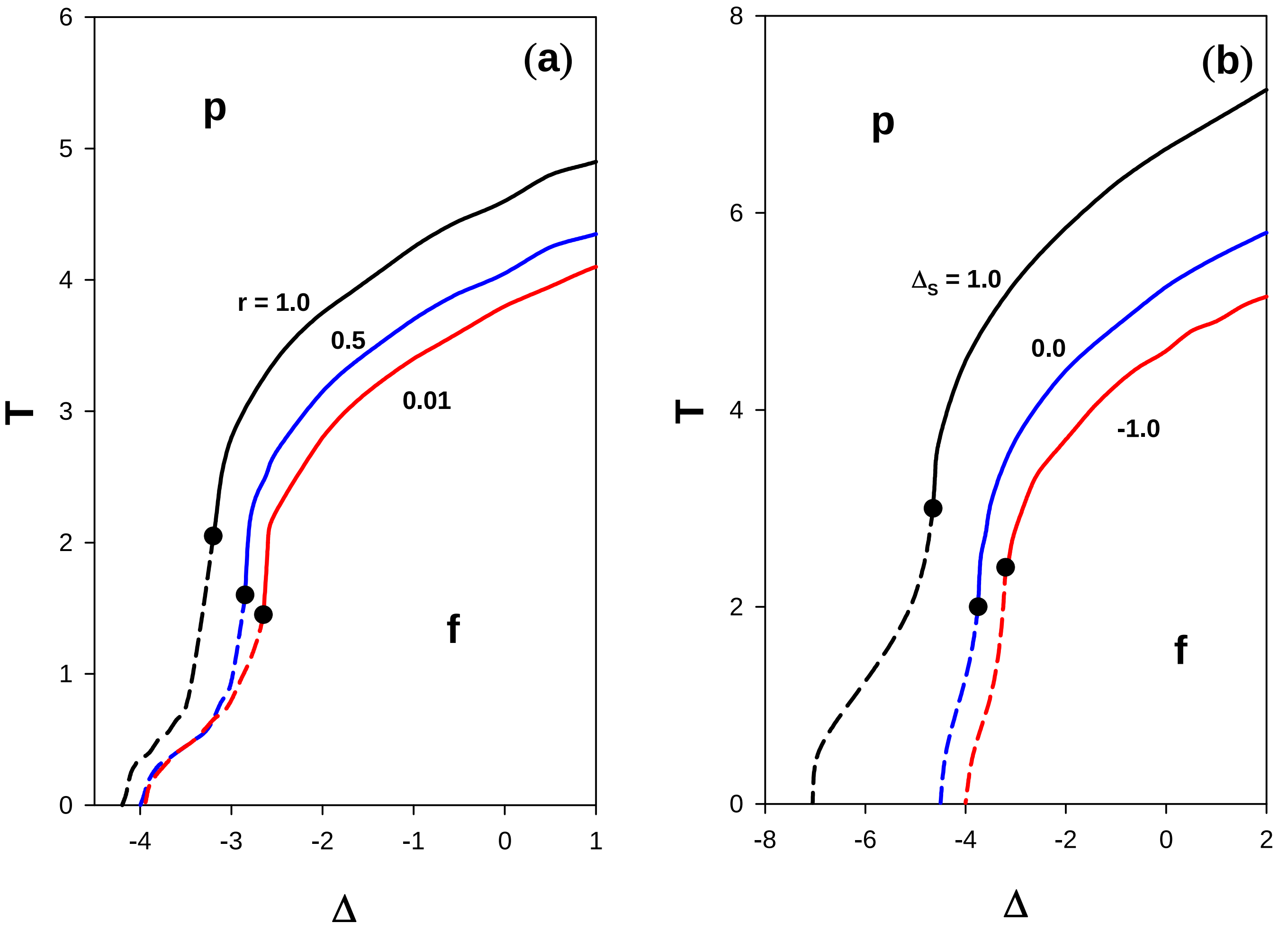

**Fig. 2**

Figure 3

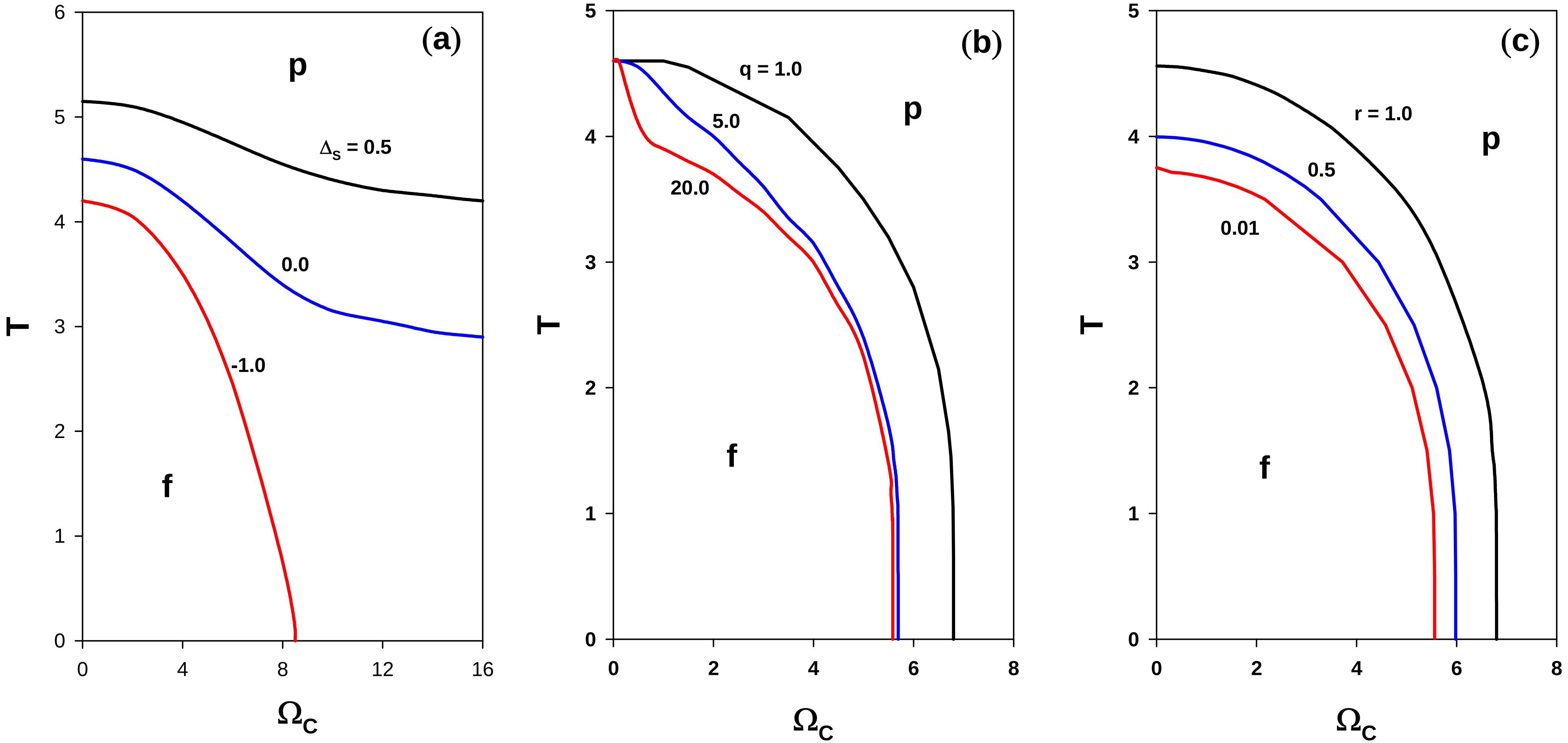

Fig. 3

Figure 4

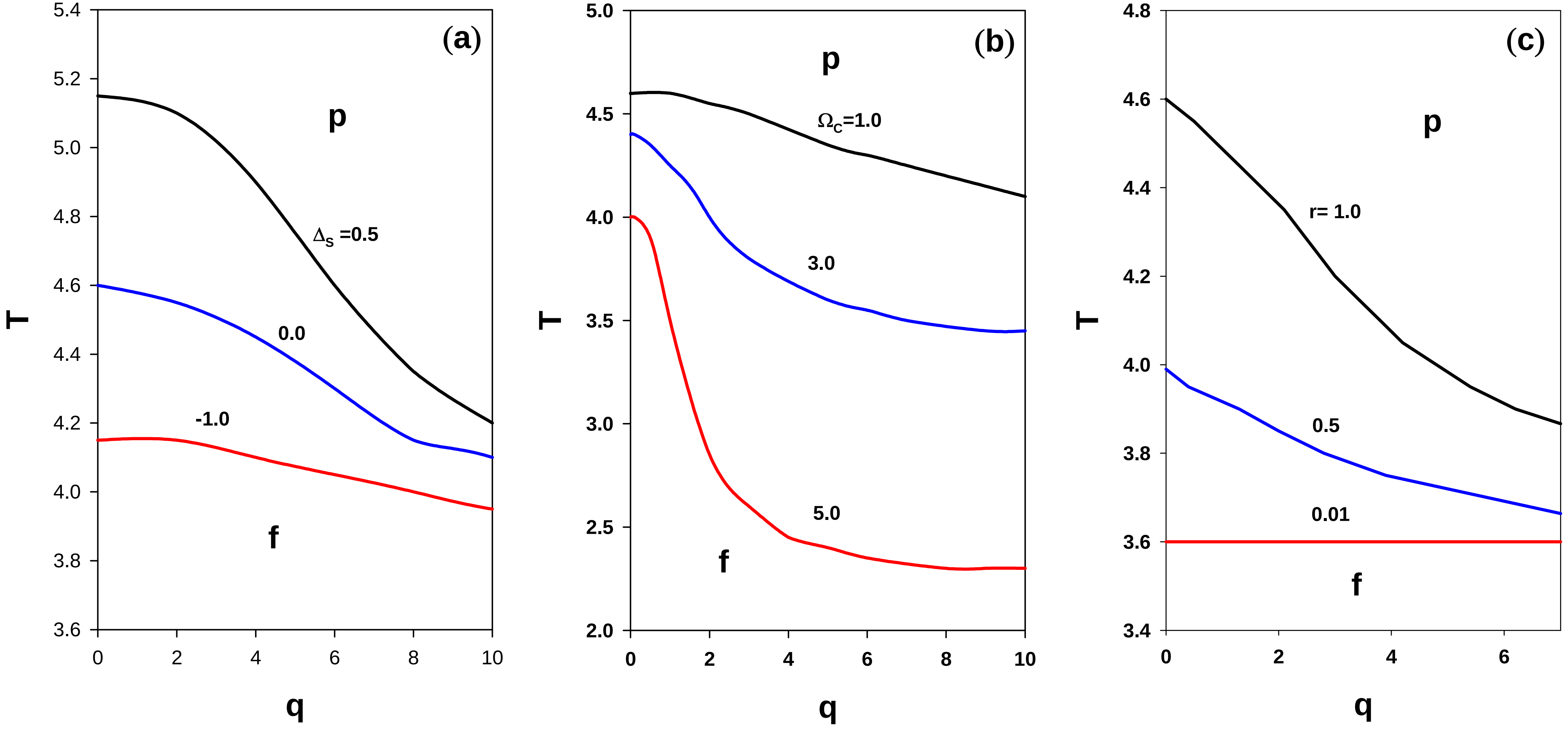

Fig. 4

# Phase diagrams of a cylindrical spin-1 Ising nanowire within the presence of the crystal field and the transverse field

Ersin Kantar[a] and Yusuf Kocakaplan[b]

[a]*Department of Physics, Erciyes University, 38039 Kayseri,Turkey*
[b]*Institute of Science, Erciyes University, 38039 Kayseri,Turkey*

## Graphical abstract

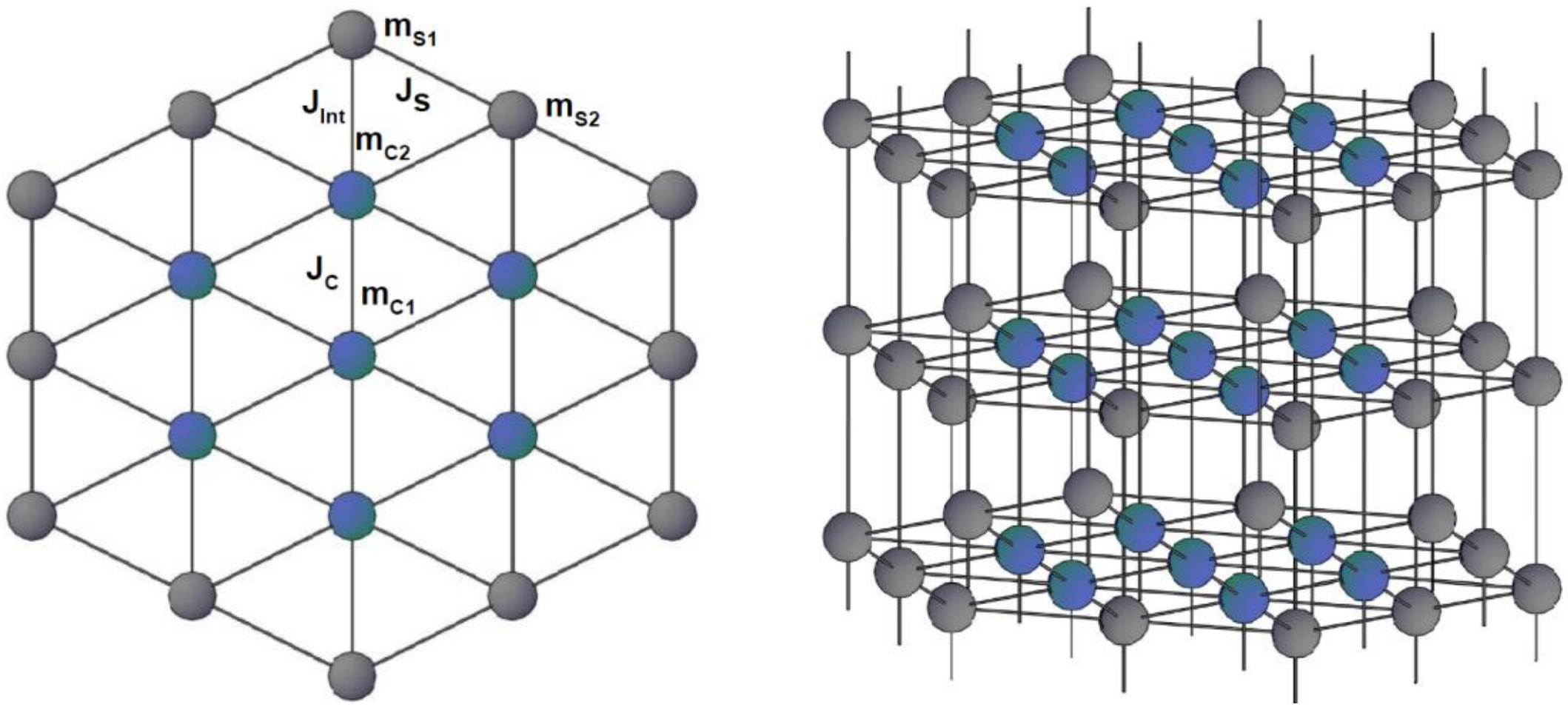